\documentclass[%
 reprint,
superscriptaddress,
 amsmath,amssymb,
 aps,
prl,
]{revtex4-1}

\usepackage{graphicx}
\usepackage{dcolumn}
\usepackage{bm}

\usepackage[caption=false]{subfig}	
\usepackage{here}
\usepackage{color}
\usepackage{ulem}

\begin{document}

\preprint{APS/123-QED}

\title{Nano-scale ultra-dense Z-pinches formation from laser-irradiated nanowire arrays}

\author{Vural Kaymak}
\email{vural.kaymak@tp1.uni-duesseldorf.de}
\affiliation{%
 Institut f\"ur Theoretische Physik, Heinrich-Heine-Universit\"at D\"usseldorf, 40225 D\"usseldorf, Germany
}%
\author{Alexander Pukhov}
\affiliation{%
 Institut f\"ur Theoretische Physik, Heinrich-Heine-Universit\"at D\"usseldorf, 40225 D\"usseldorf, Germany
}%
\author{Vyacheslav N. Shlyaptsev}%
\affiliation{Department of Electrical Computer Engineering, Colorado State University, Fort Collins, Colorado 80523, USA}
\author{Jorge J. Rocca}%
\affiliation{Department of Electrical Computer Engineering, Colorado State University, Fort Collins, Colorado 80523, USA}
\affiliation{Department of Physics, Colorado State University, Fort Collins, Colorado 80513, USA}

\date{\today}

\begin{abstract}

We show that ulta-dense Z-pinches with nanoscale dimensions can be generated by irradiating aligned nanowires with femtosecond laser pulses of relativistic intensity. Using fully three-dimensional relativistic particle-in-cell simulations we demonstrate that the laser pulse drives a forward electron current in the area around the wires. This forward current induces return current densities of $\sim$ 0.1 Giga-Amperes per $\mu$m\textsuperscript{2} through the wires.  The resulting strong, quasi-static, self-generated azimuthal magnetic field pinches the nanowires into hot plasmas with a peak electron density of $> 9\cdot 10^{24}$ cm\textsuperscript{-3}, exceeding 1000 times the critical density. Arrays of these new ultra-dense nanopinches can be expected to lead to efficient micro-fusion and other applications.

\begin{description}
\item[PACS numbers]
52.38.Dx, 52.38.Kd, 52.38.Ph
\end{description}
\end{abstract}

\pacs{Valid PACS appear here}
\maketitle


Z-pinches or Bennett pinches \cite{Bennett} are plasmas compressed by the radially inward Lorentz force resulting from an axial current and its self-generated azimuthal magnetic field. Z-pinches occur in nature, and are also routinely created in the laboratory. They have been extensively studied \cite{Haines1,Haines2}. Natural Z-pinches include interstellar filaments, solar flares and lightning bolts \cite{Severnyi, Ciardi}. In the laboratory, their ability to concentrate large amounts of energy into small volumes has attracted attention for more than half a century as potential fusion devices, as sources of intense flashes of X-rays \cite{CoverdaleXrays} and neutrons \cite{RuizNeutrons, CoverdaleNeutrons}, and to study astrophysical phenomena such as opacity at stellar-interior conditions \cite{Bailey}. Initial excitement was created by the possibility of directly driving a thermonuclear burn by applying a sufficiently large current. However, the strong magneto-hydrodynamic instabilities that develop \cite{Haines1, Kruskal, Tayler} have been a major obstacle to this and other applications. More recently the development of new techniques such as the use of exploding nanowire arrays \cite{Sanford}, and the careful pre-ionization and rapid excitation of capillary channels \cite{Rocca} have allowed for a high degree of control of these dense plasmas. These advances have made possible, for example, the creation of X-ray pulses of unsurpassed energy \cite{Stygar} and the generation of bright soft X-ray laser beams on a table-top \cite{Rocca2, Benware}.

The size of Z-pinches spans over an enormous range. While the dimension of some naturally occurring pinches are measured in astronomical units, the diameter of laboratory pinches typically ranges from millimeters to the tens of micrometers. In this letter we  demonstrate a new kind of dense Z-pinch with nanometer-scale radial dimension and ultra-high plasma density, $n_e > 9 \cdot 10^{24}$ cm\textsuperscript{-3}, which can be created in nanowires irradiated by femtosecond laser pulses of relativistic intensity.  We show that the irradiation of arrays of aligned nanowires with intense, high-contrast, ultrashort laser pulses along the nanowire axis creates a large forward current in the voids between the wires. This forward current in the vicinity of the wires gives rise to a return current through the wires to maintain quasi-neutrality. Our simulations show that when irradiated with a relativistic laser intensity of $4.95\cdot 10^{21}$ W cm\textsuperscript{-2} ($a_0 = 17$) the current density within the nanowires reaches 96 MA/$\mu$m\textsuperscript{2}. This large return current in turn generates a strong quasi-static azimuthal magnetic field surrounding the nanowire. The resulting Lorentz force compresses the nanowires into an extremely hot and dense plasma where the electron density reaches values 1000 times larger than the critical density. It should be noted that side-on irradiation of 20 $\mu$m diameter wires with intense picosecond pulses of 60-80 J energy has been reported to generate a localized MHD m = 0 instability on the nanosecond time scale, similar to that observed in discharge-driven Z-pinches, resulting in field emission from nearby objects \cite{Beg}. The electron density in the wire reached 20 times the critical density. Earlier experiments used $\lambda = 10.6 \mu$m laser pulses of nanosecond duration to generate a laser-produced return current to implode 130 $\mu$m diameter thin cylindrical liners \cite{Hauer}, and to study the ohmic heating of thin fibers \cite{Benjamin}. The new type of nanowire pinch of interest here is driven by femtosecond laser pulses of relativistic intensity and has diameters two to three orders of magnitude smaller, leading to densities that exceed by more than 1000 times the critical density.
 
The targets of interest for the formation of dense nano-scale Z-pinches are arrays of aligned nanowires. Nanowire arrays are typically defined by local solid density within the nanowire and by average densities that range from several percent to a large fraction of solid density. They have been shown to facilitate a high absorption of incident laser light \cite{Purvis}. Moreover, contrary to the case of solid target surfaces, in which the rapid formation of a critical density layer prevents the efficient coupling of laser energy deep into the material, arrays of aligned, high-aspect-ratio nanowires allow for light penetration deep inside the near-solid density nanostructure. This occurs as long as the gaps between the nanowires are free from plasma. It was recently demonstrated that irradiation of ordered nanowire arrays with femtosecond pulses of relativistic intensity can volumetrically heat dense matter to an ultra-hot dense-plasma regime defined by electron densities up 100 times the critical density, multi-keV temperatures, and extreme degrees of ionization \cite{Purvis}. Optical field ionization by the strong laser field occurs at the boundary of the nanowires where vacuum heating of electrons takes place \cite{Brunel}. Both the hot electrons that penetrate into the nanowire core and the cold electrons that form the return current cause intensive collisional ionization of the material. However, the contribution of the return current is larger due to the increased collision cross-section. Here we demonstrate that the large reverse electron current in the nanowire rods leads to a nano-scale Z-pinch effect that further increases the electron density of the hot material by up to nearly two orders of magnitude. Another important consequence of the compression of the wires is that the confinement retards the expansion of the nanowire plasma, effectively delaying the closure of the inter-wire gaps. This increases the time available to couple the laser energy deep into the material to achieve efficient volumetric heating.

Below we discuss the physics of this new nanoscale ultra-dense Z-pinch regime, presenting results of full three-dimensional (3D) relativistic particle-in-cell simulations for an array of aligned carbon nanowires of 300 nm diameter and 5$\mu$m length. The simulations were conducted using the relativistic Virtual Laser Plasma Lab (VLPL) particle-in-cell code \cite{Pukhov}. The periodicity of the array is 1$\mu$m, which corresponds to an average density of 7\% solid density. The array is irradiated by 400-nm-wavelength laser pulses of 60 fs FWHM duration and a vector potential of $a_0$ = 17, corresponding to an intensity of $4.95\cdot 10^{21}$ W cm\textsuperscript{-2}. The laser beam is modeled as a circularly polarized plane wave with a gaussian temporal profile impinging at normal incidence onto the nanowire array. The code incorporates optical field ionization and electron impact ionization, as well as binary collisions. The irradiation of the nanowires with pulses of relativistic intensity leads to rapid ionization of the material.

\begin{figure}[H]
\centering
\includegraphics[width=8.6cm]{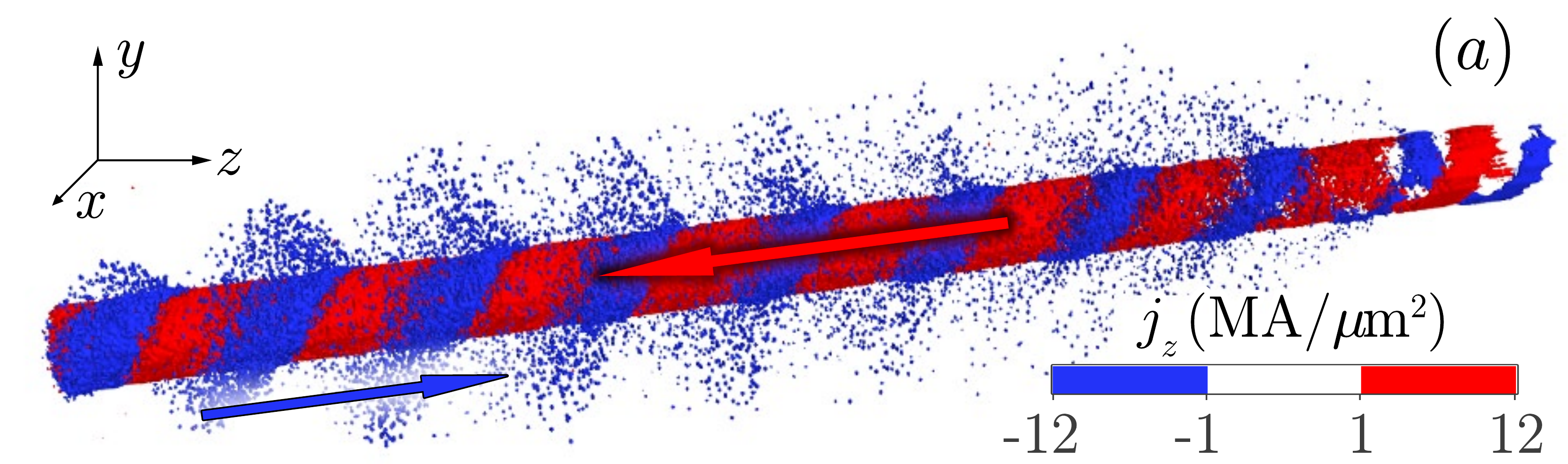}\vspace{0.25cm}
\includegraphics[width=8.6cm]{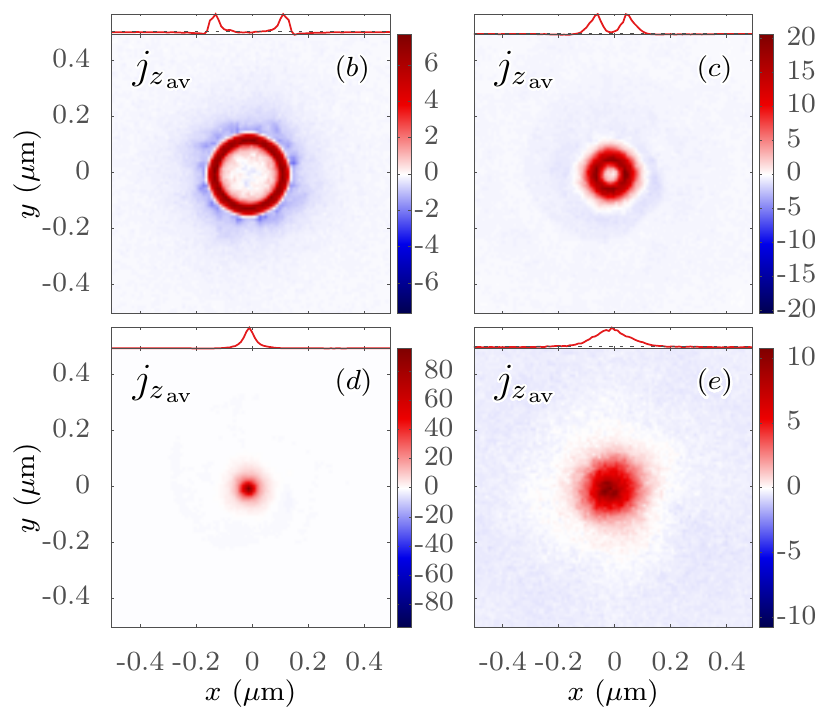}
\caption{Frame (a): Current density $j_z$ at $t = -44.5 T_0$ before arrival of the peak of the pulse to the target surface. Electrons pulled out of the wire move in the laser propagation direction (blue color, from left to right). A return current is created (red color, from right to left). Frames (b)-(e): Cross-sections of the current density distribution, $j_z$ [MA/$\mu$m\textsuperscript{2}], averaged along the wire axis in the plane perpendicular to the laser propagation axis over the narrowest nanowire segment. The times are (b) $t=-26 T_0$, (c) $t=-13 T_0$, (d) $t=-7 T_0$ and (e) $t=8 T_0$. The one dimensional cuts are taken at the center along the $x$ axis. The times of the snapshots are given with respect to the peak of the pulse reaching the target surface in units of the laser period ($T_0 = 1.33$ fs). The wires are 300 nm in diameter and are irradiated by a $\lambda = 400$ nm laser pulse of 60 fs duration with $a_0 = 17$.}
\label{a:avjx}
\end{figure}

The intense laser pulse rips off electrons along the electric field $\vec{E}$ and accelerates them forward via the $\vec{v} \times \vec{B}$ force. This forward current of electrons in the vicinity of the wires is compensated by a large return current inside the nanorod such that quasi-neutrality is maintained. Due to the circular laser polarization these currents are arranged along a spiral path. This is illustrated in Fig.~\ref{a:avjx}a, where the forward-moving electrons in the voids of the wires, defined here as a negative current, are indicated in blue, while the positive current densities corresponding to the return current of the backward flowing electrons is indicated in red. The domains of the forward and return current are spatially separated (Fig.~\ref{a:avjx}). The return current is initially located in the skin layer of the nanowire (Fig.~\ref{a:avjx}b). However, an emerging quasistatic azimuthal magnetic field caused by this current exerts a radially symmetric inward directed $\vec{j}\times\vec{B}$ force on the electrons. Consequently, the return current is radially compressed (Fig.~\ref{a:avjx}c).

This process compresses the return current to a thin filament concentrated at the nanowire axis (Fig.~\ref{a:avjx}d), starting with the segment near the tip where the laser field is the most intense. The backward-flowing electrons are thereby accelerated to an average velocity of up to $v \sim 0.23 c$ and the return current density reaches a value of  $j_z = 96$ MA/$\mu$m\textsuperscript{2} (Fig.~\ref{a:avjx}d), where $T_0$ is the laser period (1.33 fs). The corresponding total current through the cross-section is $\sim 21 I_0$, with $I_0 \approx 17$ kA the Alfv\'{e}n current.

\begin{figure}[H]
\centering
  \includegraphics[width=8.6cm]{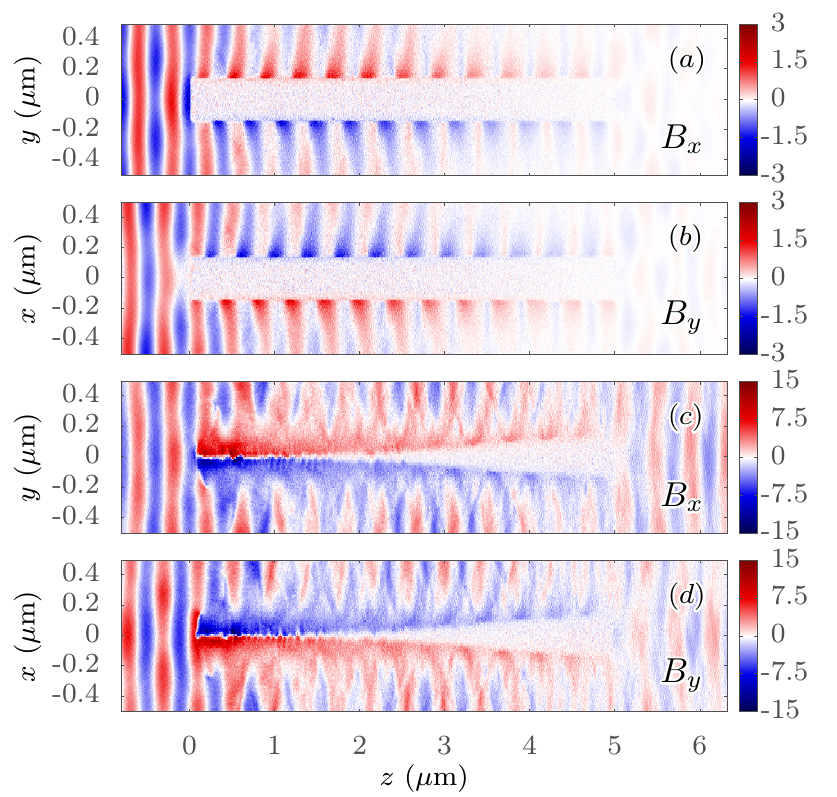}
\caption{Longitudinal cross-sections of magnetic field components $B_x$ and $B_y$ (in Gigagauss) at $t=-35 T_0$ (frames a,b) and $t=-7 T_0$ (frames c,d). Laser propagation is from left to right.}
\label{a:fields}
\end{figure}

\begin{figure}[H]
\centering
  \includegraphics[width=8.6cm]{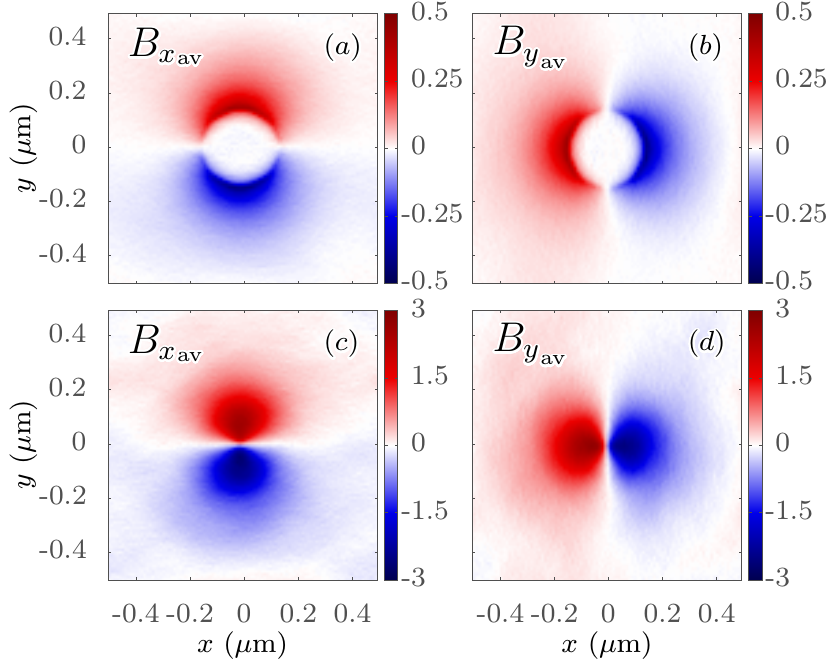} 
\caption{Quasi-static magnetic field components in Gigagauss (averages of $B_x$ and $B_y$ along the wire axis in the plane perpendicular to the laser axis) at (a-b) $t=-35 T_0$ and (c-d) $t=-7 T_0$.}
\label{a:avfields}
\end{figure}

The longitudinal cross-sections of the magnetic field components illustrated in  Fig.~\ref{a:fields} reveal the emerging quasistatic field. The laser propagation direction is from left to right. Initially this field is located at the surface of the nanowire, separated from the alternating component of the laser (Fig.~\ref{a:fields}a and b). As the laser pulse propagates deeper into the nanowire array, the quasistatic field fills the space initially occupied by the nanorod (Fig.~\ref{a:fields}c and d), surrounding the compressed wire. To separate the quasistatic magnetic field, one can average the components $B_x$ and $B_y$ of the magnetic field along the wire axis as shown in Fig.~\ref{a:avfields}. Considering the $B_x$ and $B_y$ components together, it becomes clear that an azimuthal quasistatic magnetic field surrounding the wire is present. The clockwise polarity of that field indicates that it originates from the backward flow of electrons. As can be seen in the averaged current-density profiles (Fig.~\ref{a:avjx}) and the magnetic field cross-sections (Fig.~\ref{a:fields}), the averaged transverse magnetic field component encircles a wire of a drastically reduced radius at a time close to the peak of the pulse.

\begin{figure}[H]
\centering
\includegraphics[width=8.6cm]{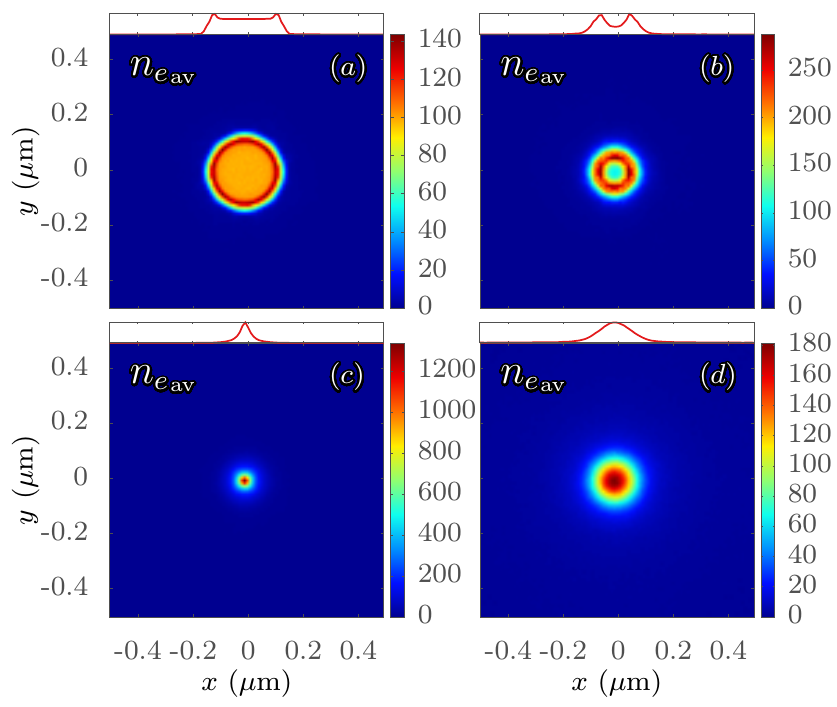}
\caption{Cross-sections of the electron number density distribution, $n_e$ (in units of the critical electron density $n_{cr} = 7 \cdot 10^{21}$ cm\textsuperscript{-3}), averaged along the wire axis in the plane perpendicular to the laser propagation axis over the narrowest nanowire segment. The frame times are (a) $t=-26 T_0$, (b) $t=-13 T_0$, (c) $t=-7 T_0$ and (d) $t=8 T_0$. The one dimensional cuts on the top are taken at the center along the $x$ axis.}
\label{a:avn0}
\end{figure}

\begin{figure}[H]
\centering
\includegraphics[width=8.6cm]{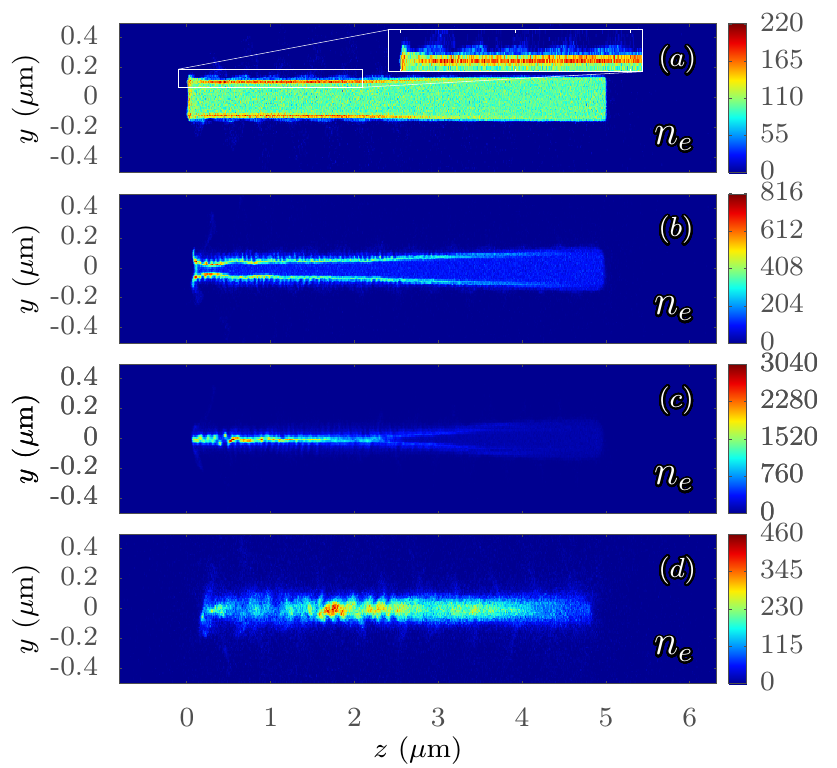}
\caption{Longitudinal cross-sections of electron particle density, $n_e$ (in units of the critical electron density $n_{cr} = 7 \cdot 10^{21}$ cm\textsuperscript{-3}), at times (a) $t=-35 T_0$, (b) $t=-13 T_0$, (c) $t=-7 T_0$ and (d) $t=8 T_0$. The inset in frame (a) shows a zoom  of the periodic structure on the surface of the nanowire plasma.}
\label{a:n0}
\end{figure}

The quasistatic magnetic field compresses the nanowire, leading to a reduction of the radius and an increase in the particle densities. These observables enable us to measure the strength of the pinch. For that purpose, we need to consider that the increase in the electron density, $n_e$, is due to two different processes: the strong optical field ionization of the carbon ions by the laser pulse and electron collisions, and the spatial compression by the pinch effect. Since all the atoms are fully ionized long before the peak of the laser pulse (at $t = -26 T_0$), any further electron density increase is attributed to the pinch effect. This process can be observed in an end-on view of the averaged electron number density (Fig.~\ref{a:avn0}a - d). A high-density ring is formed and squeezed until the average density is highly concentrated at the wire center. Since the return current, and hence the quasistatic magnetic field, have a limited lifetime, at some point the plasma pressure becomes stronger than the confining magnetic pressure giving rise to an expansion of the wire, and consequently to a decrease of the density (Fig.~\ref{a:avn0}d). Figures~\ref{a:n0}a-\ref{a:n0}d present $y$-$z$ cross-sections of the electron density of the irradiated nanowire, giving insight to the pinch effect along the longitudinal axis. Electron bunches form a periodic structure ($\sim \lambda$ in size, where $\lambda$ is the laser wavelength) on the surface of the wire (Fig.~\ref{a:n0}a). These electron bunches are pulled out from the nanowire and pushed back into it by the laser field. The high-density layer forming around the wire tip is a sign of the pinch effect kicking in around -13 $T_0$.

As the propagating laser pulse pulls out more electrons along the wire, the return current and therefore the quasistatic magnetic field become stronger. The compression by the pinch effect then leads to an ultra-high average peak density of $n_e = 1338 n_{cr} = 9.4 \cdot 10^{24}$ cm\textsuperscript{-3} (Fig.~\ref{a:avn0}c) over a length of $\sim 2.4 \mu m$ (Fig.~\ref{a:n0}c). As the wire expands the density decreases and the periodic structure of the electrons becomes visible~(Fig. \ref{a:n0}d).

\begin{figure}[H]
\centering
\includegraphics[width=8.6cm]{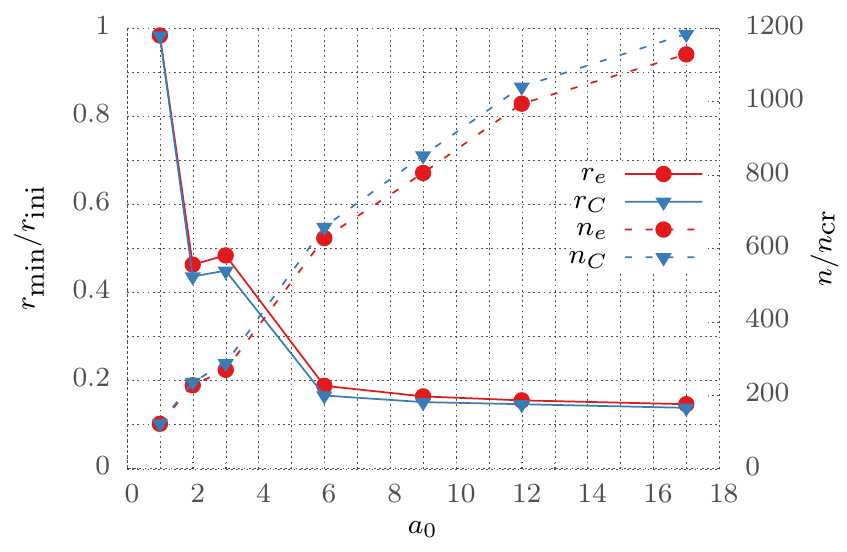}
\caption{Minimum radii normalized to the initial nanowire diameter (solid lines) and corresponding averaged particle densities of electrons and carbon ions (dashed lines) in units of the critical electron density, $n_{cr} = 7 \cdot 10^{21}$ cm\textsuperscript{-3}, for seven different laser amplitudes. The values were computed by averaging over the nanowire segment with the narrowest radius and highest density. While the minimum radius for $a_0 = 2$ is $\sim$2\% smaller than for $a_0 = 3$ (see dashed line), the average radius of the whole wire is $\sim$ 7\% larger.}
\label{a:radius}
\end{figure}

Along with the increase in particle density, the reduction of the wire radius is the main signature of the Z-pinch. Figure~\ref{a:radius} shows the minimum radii, $r_{\mbox{min}}$, as a function of the laser amplitude $a_0$ calculated by averaging the plasma column radius over the narrowest segment of the nanowire. For an ultraintense laser pulse of $a_0 = 17$ the minimum observed radii normalized by the initial wire radius are $r_{\mbox{min}}/r_{\mbox{ini}} = 14.6\%$ for the electrons and $r_{\mbox{min}}/r_{\mbox{ini}} = 13.8\%$ for the carbon ions. This slight difference results from the fact that when 60 cells are used in the transverse direction the minimum radii values are slightly below the resolution limit of $\Delta r/r_{\mbox{ini}} = 15.5\%$ with respect to the initial radius $r_{\mbox{ini}}$. For a higher number of 100 cells in the transverse direction the resolution improves to $\Delta r/r_{\mbox{ini}} = 9.4\%$, and the minimum radius at $a_0 = 17$ are computed to be $r_{\mbox{min}}/r_{\mbox{ini}} = 12.5\%$ and 12.1\% for electrons and carbon ions respectively. The minimum radii for $a_0 \geq 9$ in Fig.~\ref{a:radius} are limited by the simulation resolution. Due to the gaussian density profile of the nanowire plasma (Fig.~\ref{a:avn0}), the particle density does not scale with the square of the radius as one would expect for a uniformly cylindrical compression. The strength of the pinch effect crucially depends on the laser amplitude. A higher laser intensity will lead to a stronger and faster pinch. The pinch onset takes place between $a_0$ = 1 and $a_0$ = 2. The electron temperature at that transition point increases from 1.3 keV to 5.6 keV. The carbon ion kinetic energy increases from below 0.5 MeV when no pinch is present to 6.1 MeV when it occurs. For the highest simulated laser amplitude of $a_0$ = 17 the electron temperature reaches 80.5 keV and the carbon ions gain energies up to 92 MeV.

In conclusion, we have demonstrated the generation of ultra-dense nano-scale Z-pinches in carbon nanowires irradiated by laser pulses of relativistic intensity. These extremely dense nanoscale Z-pinches will enable new fundamental studies and a variety of applications. For example, in nanowire arrays the pinch retards the nanowire expansion, increasing the time window available for the laser pulse to deposit its energy deep into the material. This in turn facilitates enhanced volumetric heating of near-solid density material to extreme temperatures and efficient X-ray generation. Substitution of the carbon nanowire material with a deuterium-containing material such as deuterated polyethylene (CD$_2$) can be expected to lead to the generation of a very high density of deuterons accelerated to energies near the peak of the cross-section for D-D fusion reactions, leading to the efficient generation of ultrafast pulses of neutrons.

V.K. thanks John Farmer for his useful advice. This material is based upon work supported by DFG TR18, EU FP7 EUCARD-2 and the Air Force Office of Scientific Research under award number FA9560-14-10232.

\bibliography{nanopinch}

\end{document}